\documentclass[article,superscriptaddress,prd]{revtex4-2}
\usepackage[colorlinks=true, pdfstartview=FitV, linkcolor=red, citecolor=blue,
urlcolor=blue]{hyperref}
\usepackage[dvipdfmx]{graphicx}
\usepackage{amsmath,amssymb}
\usepackage{color}
\usepackage{slashed}

\renewcommand\b{\beta}

\renewcommand\t{\tau}

\newcommand\g{\gamma}

\newcommand\m{\mu}
\newcommand\n{\nu}

\newcommand\p{\pi}

\newcommand\s{\sigma}

\renewcommand\L{\Lambda}

\renewcommand\O{\Omega}

\newcommand\D{\Delta}

\newcommand{\tr}{{\rm{tr}}}

\newcommand{\Lag}{\mathcal{L}}

\newcommand{\ud}{\mathrm{d}}
\newcommand{\ue}{\mathrm{e}}

\renewcommand{\part}{{\rm part}}

\begin{abstract}
Rapid rotation may exist in physical systems such as non-central heavy ion collisions and neutron stars. Using functional renormalization group analysis of the quark-meson model, we investigate the effects of real and imaginary rotation on the chiral phase transition. Our results confirm previous studies conducted with other model calculations and shed light on the importance of boundary conditions in the infrared region of the theory.
\end{abstract}

\begin{document}

\title{Quark-meson model under rotation: A functional renormalization group study}
\author{Hao-Lei Chen}
\email{hlchen15@fudan.edu.cn}
\affiliation{Key Laboratory of Nuclear Physics and Ion-beam Application (MOE), Fudan University, Shanghai 200433, China}
\affiliation{Shanghai Research Center for Theoretical Nuclear Physics, NSFC and Fudan University, Shanghai 200438, China}
\author{Zhi-Bin Zhu}
\email{20110190016@fudan.edu.cn}
\affiliation{Physics Department and Center for Particle Physics and Field Theory, Fudan University, Shanghai 200438, China}
\author{Xu-Guang Huang}
\email{huangxuguang@fudan.edu.cn}
\affiliation{Physics Department and Center for Particle Physics and Field Theory, Fudan University, Shanghai 200438, China}
\affiliation{Key Laboratory of Nuclear Physics and Ion-beam Application (MOE), Fudan University, Shanghai 200433, China}
\affiliation{Shanghai Research Center for Theoretical Nuclear Physics, NSFC and Fudan University, Shanghai 200438, China}

\maketitle

\section{Introduction}
The phenomena induced by rotation have received intensive attention in recent years. In heavy-ion collisions, extremely strong fluid vorticity (i.e., local rotation) can be generated~\cite{Deng:2016gyh,Jiang:2016woz,Deng:2020ygd}, which can induce spin polarization of spinful particles~\cite{Liang:2004ph,Liang:2004xn,STAR:2017ckg,STAR:2022fan} and give rise to a parity-violating current known as the chiral vortical effect (CVE)~\cite{Vilenkin:1979ui,Erdmenger:2008rm,Banerjee:2008th,Son:2009tf,Landsteiner:2011cp}. The chiral phase transition of quantum chromodynamics (QCD) under rotation has also been discussed recently~\cite{Chen:2015hfc,Jiang:2016wvv,Ebihara:2016fwa,Chernodub:2016kxh,Chernodub:2017ref,Chernodub:2017mvp,Wang:2018sur,Wang:2018zrn,Wang:2019nhd,Zhang:2020jux,Sadooghi:2021upd,Chen:2022mhf,Mehr:2022tfq} using the Nambu-Jona-Lasinio (NJL) model and other models, and confinement-deconfinement transition has been studied using holographic models~\cite{Chen:2020ath,Braga:2022yfe,Yadav:2022qcl,Zhao:2022uxc} and other methods~\cite{Fujimoto:2021xix,Chen:2022smf,Mukherjee:2023qvq}. The rotation-induced meson condensate has also been investigated~\cite{Huang:2017pqe,Chen:2019tcp,Cao:2020pmm,Cao:2019ctl,Liu:2017spl,Zhang:2018ome,Nishimura:2020odq}; see~\cite{Chen:2021aiq} for a recent review. Model calculations suggest that the critical temperature decreases with increasing angular velocity for both the chiral and confinement-deconfinement transitions. However, lattice simulations based on imaginary angular velocity $\Omega_I$ show that the critical temperature of both transitions decreases with $\Omega_I$\cite{Braguta:2021jgn,Yang-Huang:2023}. Therefore, if one naively performs the analytical continuation to real rotation by $\Omega^2_I\to-\Omega^2$, the critical temperature becomes an increasing function of $\Omega$. The contradiction between model calculations and lattice simulations is puzzling and requires further investigation. Some studies suggest that the analytical continuation may be problematic~\cite{Chen:2022smf,Chernodub:2022qlz}. In~\cite{Chernodub:2020qah,Chernodub:2022veq,Chernodub:2022wsw}, an inhomogeneous confinement phase is proposed, and instanton solutions with imaginary rotation are also discussed.

The contradiction between effective model calculations and lattice results of rotation may arise from the absence of the non-perturbative gluonic effects in the model calculations. To include the gluon field, we need to start with first principle calculations, and the functional renormalization group (fRG) can be a promising tool for this purpose. The fRG is a nonperturbative method that is widely used to study the phase structure of QCD at finite temperature and density. It can not only be applied to deal with low-energy effective models (e.g., NJL model and the quark-meson model) but also to directly calculate QCD from first principles. For reviews of fRG, see~\cite{Berges:2000ew,Pawlowski:2005xe, Gies:2006wv,Dupuis:2020fhh,Fu:2022gou}. As it is challenging to deal with the gluon fields in rotating frame, we adopt the quark-meson (QM) model with fRG approach~\cite{Jungnickel:1995fp,Schaefer:1999em,Schaefer:2004en,Tripolt:2014wra,Yokota:2016tip} in this work to study the chiral phase transition under rotation. This serves as a warm-up for future treatment of QCD. When combining fRG with QM model in rotating spacetime, we find that rotation leads to a singularity when dealing with mesons. To avoid this singularity, it is necessary to impose a boundary condition, which behaves as an effective infrared cutoff. Our numerical results are in qualitative agreement with previous studies based on the NJL model. However, in our case, the rotational effect is mild, and the critical end point is not reachable due to the causality restriction. To provide a comprehensive analysis, we also discuss the case of imaginary rotation and show that a smooth analytical continuation to real rotation is possible in our study.

Throughout this paper, we use natural units $\hbar=k_B=c=1$ and the convention for Minkowski metric $\eta_{\m\n}={\rm diag}(1, -1, -1, -1)$.

\section{Solution to Klein-Gordon Equation and Dirac Equation}
We start with discussing the solutions to the equations of motion for free quarks and mesons in the rotating frame. The results will be utilized in subsequent sections. To describe a system undergoing global rotation, it is convenient to go into the co-rotating frame. This can be achieved by adopting the following metric tensor:
\begin{equation}\label{eq:metric}
	g_{\m\n}=\left(
	   \begin{array}{cccc}
		 1-r^2\O^2 &\O y& -\O x& 0 \\
		 \O y & -1 & 0 & 0 \\
		 -\O x & 0 & -1 & 0 \\
		 0 & 0 & 0 & -1 \\
	   \end{array}
	 \right),
  \end{equation}
where $\Omega$ is a constant angular velocity and $r^2=x^2+y^2$. Here, we assume that the rotation is along the $z$ direction. The solutions to the Klein-Gordon equation and Dirac equation under rotation have been extensively discussed in literature (see e.g.~\cite{Ambrus:2014uqa,Ambrus:2015lfr,Mameda:2015ria,Ebihara:2016fwa}). Therefore, we present only a summary of the main results, omitting the detailed calculations.

In the rotating spacetime with metric~(\ref{eq:metric}), the Klein-Gordon equation, in the cylindrical coordinate, can be written as
\begin{equation}\label{eq:KG}
	[(\partial_t-\Omega\partial_\theta)^2-\partial_r^2-\frac{1}{r}\partial_r-\frac{1}{r^2}\partial_\theta^2-\partial^2_z+m^2]\phi=0.
\end{equation}
Since we have to preserve the causality condition $\Omega r\leq 1$, appropriate boundary condition should be imposed. Here we choose the Dirichlet boundary condition for scalar field, $\phi(r=R)=0$, with $R$ the radius of the system. Then the solution to Eq.~(\ref{eq:KG}) reads
\begin{equation}
	\phi=\frac{1}{N_{l,i}^2}e^{-i(\varepsilon-\Omega l)t+il\theta+i p_z z}J_l(p_{l,i}r),
\end{equation}
where $l$ is the quantum number of angular momentum and $p_{l,i}$ is the discretized transverse momentum which is related to the $i$-th root of the Bessel function $J_l$,
\begin{equation}
	J_l(p_{l,i}R)=0,
\end{equation}
and the normalization fractor $N_{l,i}$ is given by
\begin{equation}
	N_{l,i}^2=\frac{R^2}{2}[J_{l+1}(p_{l,i}R)]^2.
\end{equation}

Next, the Dirac equation in the rotating spacetime with metric~(\ref{eq:metric}) is
\begin{equation}
	[\gamma^0(i\partial_t+\Omega {\hat J}_z)+i\gamma^i\partial_i-m]\psi=0,
\end{equation}
where $\hat{J}_z=(-i{\vec r}\times\nabla)_z+\sigma_z/2$ is the angular momentum operator. The particle solution reads~\cite{Ebihara:2016fwa} 
\begin{equation}\label{eq:metric2}
	u_+=\frac{\ue^{-i(\varepsilon-\O j)+ip_zz}}{\sqrt{\varepsilon+m}}\left(
	   \begin{array}{c}
		 (\varepsilon +m)\phi_l \\
		 0  \\
		 p_z\phi_l  \\
		 i\tilde p_{l,i}\varphi_l \\
	   \end{array}
	 \right),\quad
	u_-=\frac{\ue^{-i(\varepsilon-\O j)+ip_zz}}{\sqrt{\varepsilon+m}}\left(
	   \begin{array}{c}
		0 \\
		  (\varepsilon +m)\phi_l  \\
		 - i\tilde p_{l,i}\varphi_l  \\
		 - p_z\phi_l\\
	   \end{array}
	 \right)
  \end{equation}
where $\tilde p_{l,i}$ stands for the discretized transverse momentum for fermions which is different from that of scalar bosons, $\phi_l=\ue^{il\theta}J_l(\tilde p_{l,i}r)$, $\varphi_l=\ue^{i(l+1)\theta}J_{l+1}(\tilde p_{l,i}r)$, and $\varepsilon =\sqrt{p_{l,i}^2+p_z^2+m^2}$. The anti-particle solution can be easily obtained by charge conjugation, $v_\pm=i\g^2u_\pm^*$. 
Similar to the Klein-Gordon case, appropriate boundary condition must be imposed. Here we choose the following boundary condition:
\begin{equation}
	\left\{
             \begin{array}{ll}
             &J_{l}(\tilde p_{l,i}R)=0,\quad l\geq 0,  \\
             &J_{-l-1}(\tilde p_{l,i}R)=0,\quad l< 0 . 
             \end{array}
\right.
\end{equation}
This choice leads to
\begin{equation}
	\int^{2\pi}_0\ud\theta\bar\psi\g^r\psi{\Big|}_{r=R}=0,
\end{equation}
i.e., there is no net current flowing into or out at the cylindrical boundary. The normalization factor for fermion is given by
\begin{equation}
	\tilde N_{l,i}=\left\{
             \begin{array}{ll}
             &\frac{R^2}{2}[J_{l+1}(p_{l,i}R)]^2,\quad l\geq 0,  \\
             &\frac{R^2}{2}[J_{l}(p_{l,i}R)]^2,\quad l< 0 .  \\
             \end{array}
\right   .
\end{equation}

\section{Quark-meson model and fRG flow equation}
Quark-meson (QM) model is a low-energy effective model for QCD which is widely used to discuss chiral symmetry breaking in different environments, such as at finite temperature and densities~\cite{Jungnickel:1995fp,Schaefer:1999em,Schaefer:2004en,Tripolt:2014wra,Yokota:2016tip}. In this work, we adopt the two flavor QM model in a rotating frame to study the rotational effects on chiral symmetry breaking. The QM model Lagrangian in Euclidean spacetime with rotation reads
\begin{equation}\label{eq:QMLag}
	\begin{split}
		\Lag=&\phi[-(-\partial_\tau+\Omega \hat L_z)^2-\nabla^2]\phi+U(\phi)\\
		     &+\bar q[\gamma^0(\partial_\tau-\Omega \hat J_z)-i\gamma^i\partial_i+g(\sigma+i\vec\pi\cdot\vec\tau\gamma^5)]	q,
	\end{split}
\end{equation}
where $\tau$ is the imaginary time, $\hat{L}_z=(-i{\vec r}\times\nabla)_z$ is the orbital angular momentum operator, and $\hat{J}_z={\hat L}_z+\hat{S}_z$ is the total angular momentum operator for quarks with $\hat{S}_z=\sigma_z/2$ the spin. The first line corresponds to the meson sector with the meson field defined as $\phi=(\sigma,\vec\pi)$ and $U(\phi)$ is the potential term for meson field. In the following we will use the abbreviation $M=\sigma+i\vec\pi\cdot\vec\tau\gamma^5$ and choose the potential term as
\begin{equation}
	\begin{split}
	U(\phi)&=\frac{m^2}{2}\phi^2+\frac{\lambda}{4}\phi^4-c\sigma,
	\end{split}
\end{equation}
where $c$ is the explicit symmetry-breaking term which gives pions a finite mass. The second line in Eq.~(\ref{eq:QMLag}) is fermion sector with the quark field $q=(u,d)$.

The fRG equation (the Wetterich equation) reads~\cite{Wetterich:1992yh}
\begin{equation}\label{eq:Wetterich}
	\partial_k\Gamma_k=\frac{1}{2}\tr(G_{\phi,k}\partial_kR_{\phi,k})-\tr(G_{q,k}\partial_kR_{q,k}),
\end{equation}
which describes the evolution of the scale-dependent effective action $\Gamma_k$ from the initial UV
scale ($k = \Lambda$) to the IR limit ($k = 0$). Here,
$R_{\phi,k}$ and $R_{q,k}$ are cutoff functions (regulators) for mesons and quarks, respectively, while $G_{\phi,k}$ and $G_{q,k}$ are full propagators of mesons and quarks. This fRG equation is exact and in general is very challenging to solve. To proceed, we thus will use the local potential approximation (LPA)~\cite{Schaefer:2004en}, in which the RG-scale dependence only enters the effective potential $U_k(\phi)$.

Let us first consider the meson sector. The regulator
$R_{\phi,k}$ that suppresses the fluctuations with momentum smaller than
the scale $k$ are chosen as the optimized regulator (Litim regulator)~\cite{Litim:2000ci,Litim:2001up} which, in the momentum space, reads
\begin{equation}
	R_{\phi,k}=(k^2-p^2)\theta(k^2-p^2),
\end{equation}
where $p^2=p_t^2+p_z^2$ with $p_t=\sqrt{p_x^2+p_y^2}$ the transverse momentum. Then the improved effective action of mesons is
\begin{equation}
\begin{split}
	\Gamma_k^B
	&=\frac{1}{2}\ln \det[-(-\partial_\tau+\Omega \hat L_z)^2-\nabla^2+\frac{\partial^2U}{\partial\phi_i\partial\phi_j}]\\
	&=\frac{1}{2}\int d^4x_E T\sum_n\int\frac{dp_z}{2\pi}\frac{1}{2\pi}\sum_{l,i}\frac{1}{N_{l,i}^2}{\tr}\ln[-(i\omega_n+\Omega l)^2+p_{l,i}^2+p_z^2+R_{\phi,k}+\frac{\partial^2U}{\partial\phi_i\partial\phi_j}]J_l(p_{l,i}r)^2,
\end{split}
\end{equation}
where $\omega_n=2\pi n T$ is the Matsubara frequency for mesons. Taking the derivatives with respect to the scale $k$, we obtain
\begin{equation}\label{eq:bpart}
	\begin{split}
		\partial_k\Gamma^B_k=\frac{1}{2}\int d^4x_E T\sum_n\int\frac{dp_z}{2\pi}\frac{1}{2\pi}\sum_{l,i}\frac{1}{N_{l,i}^2}{\tr}\frac{2k\theta(k^2-p^2)}{-(i\omega_n+\Omega l)^2+k^2+\frac{\partial^2U}{\partial\phi_i\partial\phi_j}}J_l(p_{l,i}r)^2.
	\end{split}
\end{equation}
If we define the modified propagator as
\begin{equation}
	\hat G_{\phi,k}^{-1}=-(-\partial_\tau+\Omega \hat L_z)^2-\nabla^2+\hat R_{\phi,k}+\frac{\partial^2U}{\partial\phi_i\partial\phi_j},
\end{equation}
where $\hat R_{\phi,k}$ is the operator form of $R_{\phi_,k}$, then Eq.~(\ref{eq:bpart}) gives exactly the meson sector of Eq.~(\ref{eq:Wetterich}).

For the quark sector, since the expression in momentum space is cumbersome, we define the regulator in operator form
\begin{equation}
	\hat R_{q,k}=-i\gamma^i\partial_i(\frac{k}{\sqrt{-\nabla^2}}-1)\theta(k^2+\nabla^2).
\end{equation}
Then the effective action for fermion is given by
\begin{equation}
	\begin{split}
	\Gamma^F_k&=-\frac{1}{2}\ln \det (D_k\gamma^5 D_k^\dagger\gamma^5)\\
	&=-\frac{1}{2}\tr\ln[-(\partial_\tau-\Omega \hat J_z)^2-\nabla^2+\hat R_{q,k}^2+gMM^\dagger]\\
	&=-\frac{1}{2}\int d^4x_E T\sum_n\int\frac{dp_z}{2\pi}\frac{1}{2\pi}\sum_{l,i}\frac{1}{N_{l,i}^2}2N_cN_f\ln[(\nu_n+i\Omega j)^2+\tilde p^2+R_{\phi,k}+g^2\phi^2][J_l(\tilde p_{l,i}r)^2+J_{l+1}(\tilde p_{l,i}r)^2],
	\end{split}
\end{equation}
where the angular momentum quantum  number $j=l+1/2$, $\nu_n=(2n+1)\pi T$ is the Matsubara frequency for quarks, and we have used the fact that $\hat R_{q,k}^2$ will become $R_{\phi,k}$ in momentum space. Taking derivatives with respect to $k$, we obtain
\begin{equation}
	\partial_k \Gamma_k^F=-\frac{1}{2}\int d^4x_E T\sum_n\int\frac{dp_z}{2\pi}\frac{1}{2\pi}\sum_{l,i}\frac{1}{\tilde N_{l,i}^2}2N_cN_f\frac{\partial_k R_{\phi,k}}{(\nu_n+i\Omega j)^2+\tilde p^2+R_{\phi,k}+g^2\phi^2}[J_l(\tilde p_{l,i}r)^2+J_{l+1}(\tilde p_{l,i}r)^2].
\end{equation}
Similar to the meson sector, we can also re-write this equation as the fermion sector in Eq.~(\ref{eq:Wetterich}) by defining
\begin{equation}
	\hat G_{\psi,k}^{-1}=\g^0(-\partial_\tau+\Omega \hat J_z)-\g^i\partial_i+\hat R_{q,k}+g\phi.
\end{equation}

Finally, after summing over Matsubara frequencies, the flow equation of potential becomes
\begin{equation}\label{eq:flow_avg}
\begin{split}
\partial_kU_k=&\frac{1}{\beta V}(\partial_k \Gamma_k^B+\partial_k \Gamma_k^F)\\
=&\frac{1}{\beta V}\int d^4x_E \frac{1}{(2\pi)^2}\Big\{\sum_{l,i}\frac{1}{N_{l,i}^2}{\tr}\frac{k\sqrt{k^2-p_{l,i}^2}}{\varepsilon_\phi}\frac{1}{2}[\coth\frac{\beta(\varepsilon_\phi+\Omega l)}{2}+\coth\frac{\beta(\varepsilon_\phi-\Omega l)}{2}]J_l(p_{l,i}r)^2\\
&-\sum_{l,i}\frac{1}{\tilde N_{l,i}^2}2N_cN_f\frac{k\sqrt{k^2-\tilde p_{l,i}^2}}{\varepsilon_q}\frac{1}{2}[\tanh\frac{\beta(\varepsilon_q+\Omega j)}{2}+\tanh\frac{\beta(\varepsilon_q-\Omega j)}{2}][J_l(\tilde p_{l,i}r)^2+J_{l+1}(\tilde p_{l,i}r)^2]\Big\},
\end{split}
\end{equation}
where $N_f=2$, $N_c=3$, and the energies are defined as
\begin{eqnarray}
	\varepsilon_\sigma&=&\sqrt{k^2+2\bar U'+4\rho\bar U''}=\sqrt{k^2+\partial_\sigma^2 \bar U}, \\
	\varepsilon_\pi&=&\sqrt{k^2+2\bar U'}=\sqrt{k^2+\partial_\sigma\bar U/\sigma},\\
	\varepsilon_q&=&\sqrt{k^2+g^2\rho},
\end{eqnarray}
where $\rho=\phi^2=\sigma^2$, the prime denotes the derivative with respect to $\rho$, the symmetric potential $\bar U=U+c\sqrt{\rho}$, and the scale dependent meson masses and quark mass are
\begin{eqnarray}
	m^2_\sigma&=&2\bar U'+4\rho\bar U'',\\
	m^2_\pi&=&2\bar U',\\
	m^2_q&=&g^2\rho.
\end{eqnarray}

Since we are interested in the local chiral condensate, we will treat meson field $\phi$ as a function of radius $r$ and we will apply local density approximation, i.e. assuming $\partial_r \rho(r)\ll \rho(r)$. Then we can write down a local version of Eq.~(\ref{eq:flow_avg}) at certain radius $r$ as
\begin{equation}\label{eq:FlowEqRot}
\begin{split}
\partial_kU_k(r)=&\frac{1}{(2\pi)^2}\Big\{\sum_{l,i}\frac{1}{N_{l,i}^2}{\tr}\frac{k\sqrt{k^2-p_{l,i}^2}}{\varepsilon_\phi}\frac{1}{2}[\coth\frac{\beta(\varepsilon_\phi+\Omega l)}{2}+\coth\frac{\beta(\varepsilon_\phi-\Omega l)}{2}]J_l(p_{l,i}r)^2\theta(k^2-p_{l,i}^2)\\
&-\sum_{l,i}\frac{1}{\tilde N_{l,i}^2}2N_cN_f\frac{k\sqrt{k^2-\tilde p_{l,i}^2}}{\varepsilon_q}\frac{1}{2}[\tanh\frac{\beta(\varepsilon_q+\Omega j)}{2}+\tanh\frac{\beta(\varepsilon_q-\Omega j)}{2}][J_l(\tilde p_{l,i}r)^2+J_{l+1}(\tilde p_{l,i}r)^2]\theta(k^2-\tilde p_{l,i}^2)\Big\}.
\end{split}
\end{equation}
Before proceeding to numerical calculations, let us compare the above flow equation with the one in a finite density system~\cite{Schaefer:2004en} (see Appendix \ref{sec:unbound}). The main difference is that rotation also affects mesons, which leads to an effective chemical potential $\Omega l$ for meson modes with angular momentum quantum number $l$. This effective chemical potential may cause a singularity if we do not take the boundary condition into account. Without boundary conditions, $p_{l,i}$ becomes continuous and takes values in the range $[0,\infty)$, thus for modes with small transverse momentum, we will always reach the singularity when evolving the flow equation to small $k$. On the other hand, due to the boundary condition, we have $p_{l,i}\pm\Omega l>0$ with $\Omega R\leq1$ \cite{Vilenkin:1980zv,Ebihara:2016fwa}. Thus, the step function in the first line of Eq.~(\ref{eq:FlowEqRot}) ensures that we will not suffer from a singularity in the flow equation, because the contribution from high angular momentum modes is suppressed at small energy scale $k$, and we always have $\varepsilon_\phi\pm\Omega l>0$. Therefore, we can conclude that the boundary condition naturally provides an effective infrared cutoff that is dependent on $l$.

Based on the previous discussion, we can propose an approximation scheme to simplify the calculation. The boundary condition acts as an effective infrared cutoff $\Lambda_{IR}(l)=p_{l,0}>\Omega l$ to prevent the appearance of singularity. We can incorporate this constraint by introducing a cutoff function for the running scale. Specifically, we add an infrared cutoff $|\Omega l|$ for the mode with angular momentum quantum number $l$, and modify the flow equation as follows:
\begin{equation}
\begin{split}
\partial_kU_k(r)=&\frac{1}{(2\pi)^2}\Big\{\sum_{l}\int{p_tdp_t}{\tr}\frac{k\sqrt{k^2-p_t^2}}{\varepsilon_\phi}\frac{1}{2}[\coth\frac{\beta(\varepsilon_\phi+\Omega l)}{2}+\coth\frac{\beta(\varepsilon_\phi-\Omega l)}{2}]J_l(p_t r)^2\theta(k^2-p_t^2)\theta(k-\Omega|l|)\\
&-2N_cN_f\sum_{l}\int{p_tdp_t}\frac{k\sqrt{k^2-p_t^2}}{\varepsilon_q}\frac{1}{2}[\tanh\frac{\beta(\varepsilon_q+\Omega j)}{2}+\tanh\frac{\beta(\varepsilon_q-\Omega j)}{2}]\\ &\quad\times[J_l(p_tr)^2+J_{l+1}( p_tr)^2]\theta(k^2-p_t^2)\theta(k-\Omega|j|)\Big\},
\end{split}
\end{equation}
where, the step function $\theta(k-\Omega|l|)$ for mesons and $\theta(k-\Omega|j|)$ for quarks ensure that the modes with large angular momentum quantum numbers $l$ and $j$ will not contribute to the flow equation, consistent with the presence of boundary conditions. Although the fermion sector does not suffer from singularities, we still add the cutoff function to make it consistent with the meson sector. Using this approximation, we can perform the integration on the transverse momentum $p_t$ explicitly, resulting in a simplified form for the flow equation:
\begin{equation}
\begin{split}
\partial_kU_k(r)=&\frac{1}{(2\pi)^2}\Big\{\sum_{l}{\tr}\frac{k}{2\varepsilon_\phi}[\coth\frac{\beta(\varepsilon_\phi+\Omega l)}{2}+\coth\frac{\beta(\varepsilon_\phi-\Omega l)}{2}]F_l(k,r)\theta(k-\Omega|l|)\\
&-N_cN_f\sum_{l}\frac{k}{\varepsilon_q}[\tanh\frac{\beta(\varepsilon_q+\Omega j)}{2}+\tanh\frac{\beta(\varepsilon_q-\Omega j)}{2}][F_l(k,r)+F_{l+1}(k,r)]\theta(k-\Omega|j|)\Big\},
\end{split}
\end{equation}
where we define
\begin{equation}
\begin{split}
F_l(k,r)\equiv\int^k_0 p_tdp_t \sqrt{k^2-p_t^2}J_l^2(p_tr)=\frac{1}{4}k^3(kr)^{2l}\frac{\Gamma(l+1/2)}{\Gamma(l+5/2)\Gamma(2l+1)}         {}_1F_2(l+1/2;l+5/2,2l+1;-k^2r^2).
\end{split}
\end{equation}
We have checked that this approximation scheme can work very well when the focus is on the physics within a region away from the boundary ($r \lesssim 0.9R$).

\section{Numerical results from the fRG flow equation}
There are mainly two methods to solve the fRG flow equation. The first one is the grid method, which discretizes the effective potential in field space and numerically evolves these grids with the flow equation. The second method is a Taylor expansion about a certain point of the effective potential and evolves the expansion coefficients. We have tested that the Taylor expansion method does not work well at non-zero rotation due to numerical instability (i.e., the results from different orders of the expansion do not converge), especially near the boundary. Therefore, in this work, we will apply the grid method to solve the flow equation. The initial condition at the UV limit is chosen as the bare form:
\begin{equation}
	U_\Lambda=\frac{m_\Lambda^2}{2}\phi^2+\frac{\lambda_\Lambda}{4}\phi^4-c\sigma,
\end{equation}
with parameters chosen as~\cite{Tripolt:2014wra,Yokota:2016tip}
\begin{equation}\label{eq:para1}
	\begin{split}
	m_\Lambda&=0.794\Lambda, \\
	\lambda_\Lambda&=2, \\
	c&=0.00175\Lambda^{-3},
	\end{split}
\end{equation}
where the cutoff is chosen as $\Lambda=1$ GeV. The chiral condensate $\langle\sigma\rangle$ is obtained by locating the minimum of the effective potential $U_{k=0}$. By using the initial conditions, we can reproduce the physical quark mass $m_q=300$ MeV, pion mass $m_\pi=137$ MeV, and pion decay constant $f_\pi=93$ MeV at $r=0$ with $T=0$ and $\Omega=0$~\cite{Tripolt:2014wra,Yokota:2016tip}. The system size is chosen as $R=100$ GeV$^{-1}$.

\begin{figure}[t]
\begin{minipage}[t]{0.5\linewidth}
\includegraphics[scale=0.6]{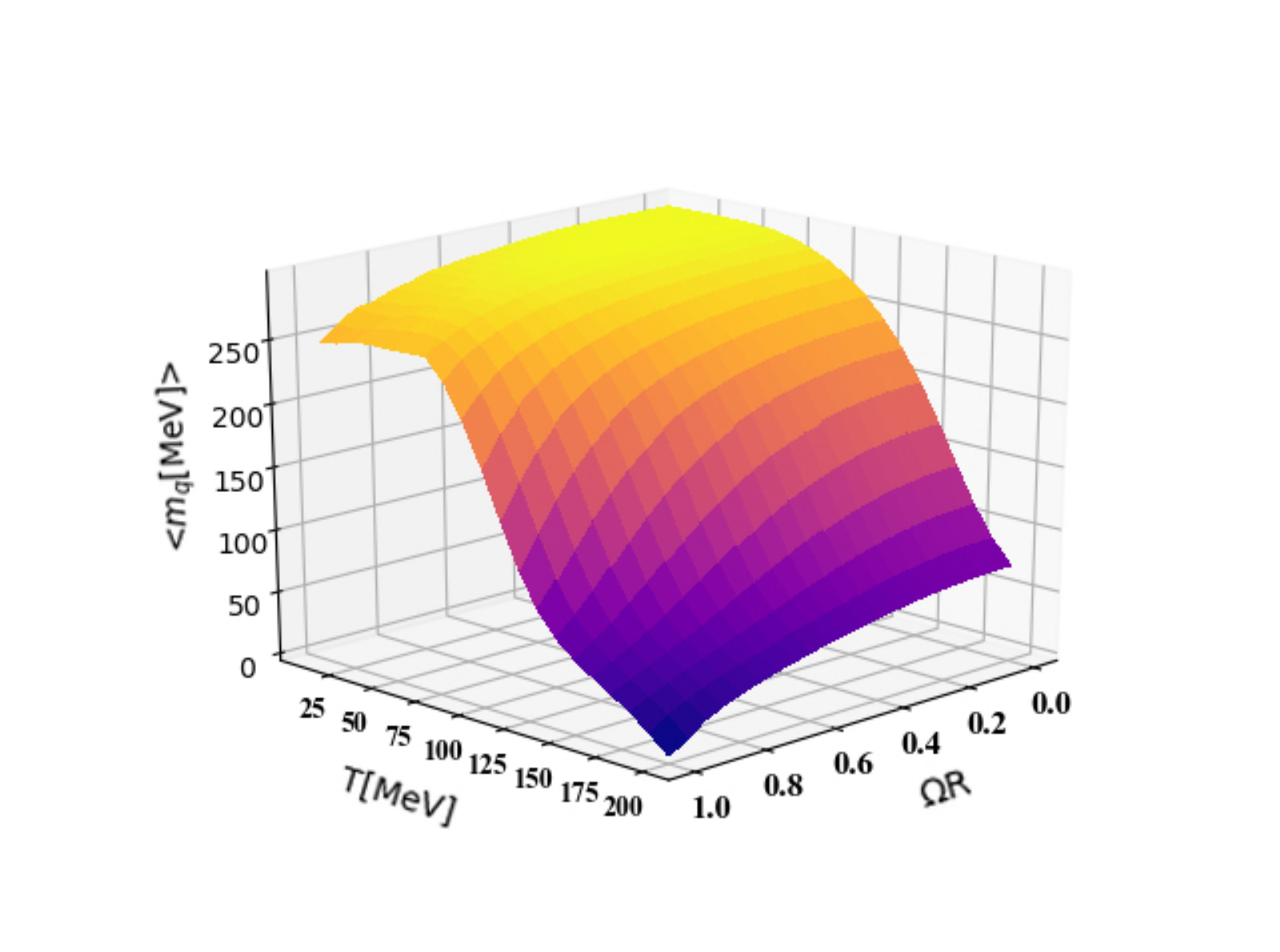}
    \caption{The quark mass $m_q$ as a function of $\Omega$ and $T$ at $r=0.9R$ in QM model.}
    \label{fig:mq_T_Omega}
\end{minipage}%
\hfill
\begin{minipage}[t]{0.45\linewidth}
\includegraphics[scale=0.5]{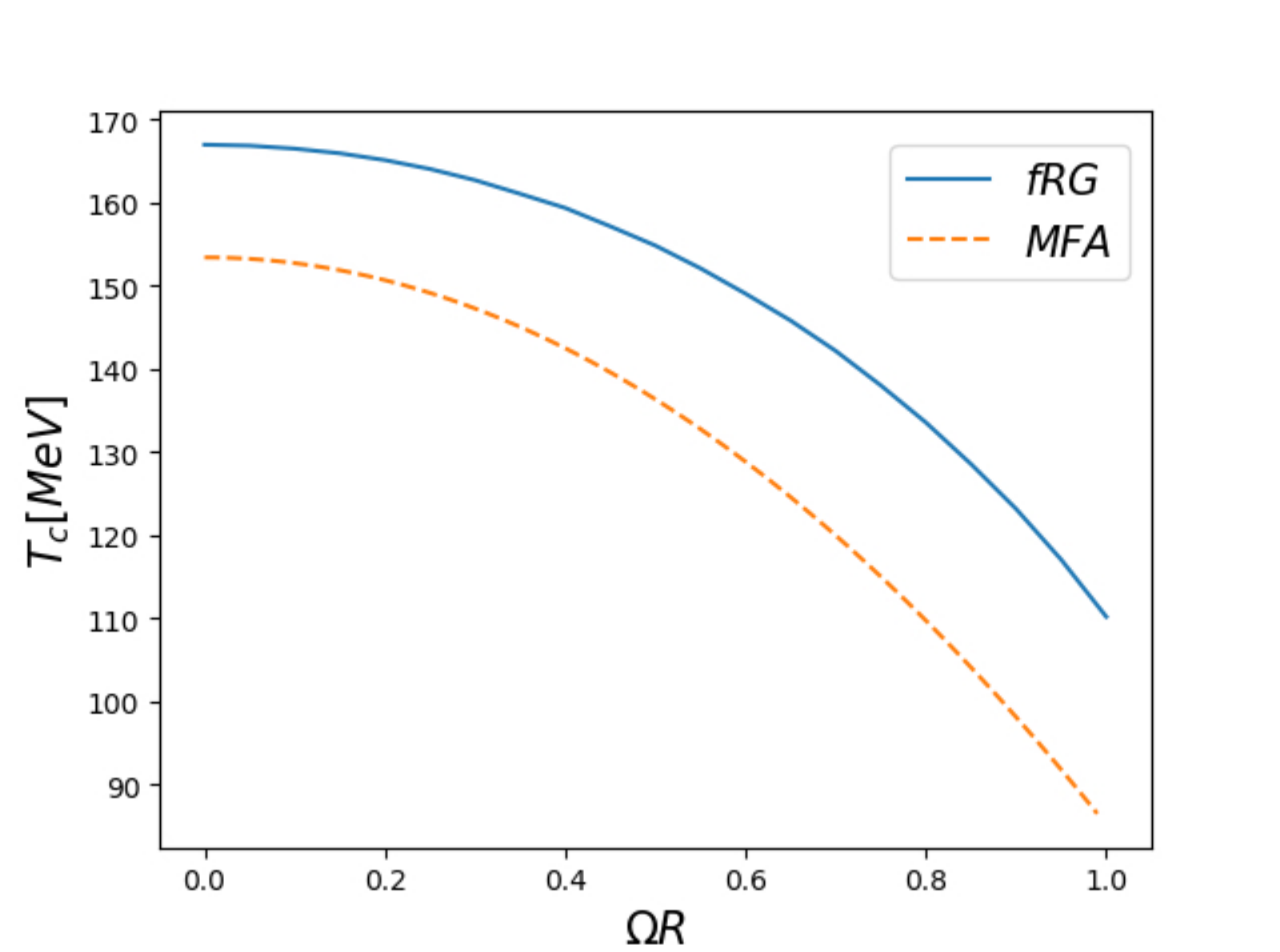}
 \caption{The pesudo-critical temperature $T_c$ as a function of $\Omega$ at $r=0.9R$ from fRG and MFA in QM model. }
    \label{fig:Tc_Omega}
\end{minipage}
\end{figure}

By solving the flow equation for the effective potential, we obtain the dependence of the quark mass on temperature $T$ and angular velocity $\Omega$, as shown in Fig.~\ref{fig:mq_T_Omega}. At low temperatures, the rotational effect is almost invisible, since ``the vacuum does not rotate", as discussed in previous works~\cite{Vilenkin:1980zv,Ebihara:2016fwa,Wang:2018zrn,Chen:2017xrj}. As the temperature increases, the rotational effect becomes visible and we can observe the suppression of chiral condensate by rotation. We can define a pseudo-critical temperature $T_c$ as the temperature where maximum of the chiral susceptibility is reached, which in our case is $1/m_\sigma^2$. The pseudo-critical temperature as a function of $\Omega R$ at $r=0.9R$ is shown in Fig.~\ref{fig:Tc_Omega}. We observe that the pseudo-critical temperature monotonically decreases with increasing rotation $\Omega$.

\begin{figure}[t]
\begin{minipage}[t]{0.45\linewidth}
\includegraphics[width=1\columnwidth]{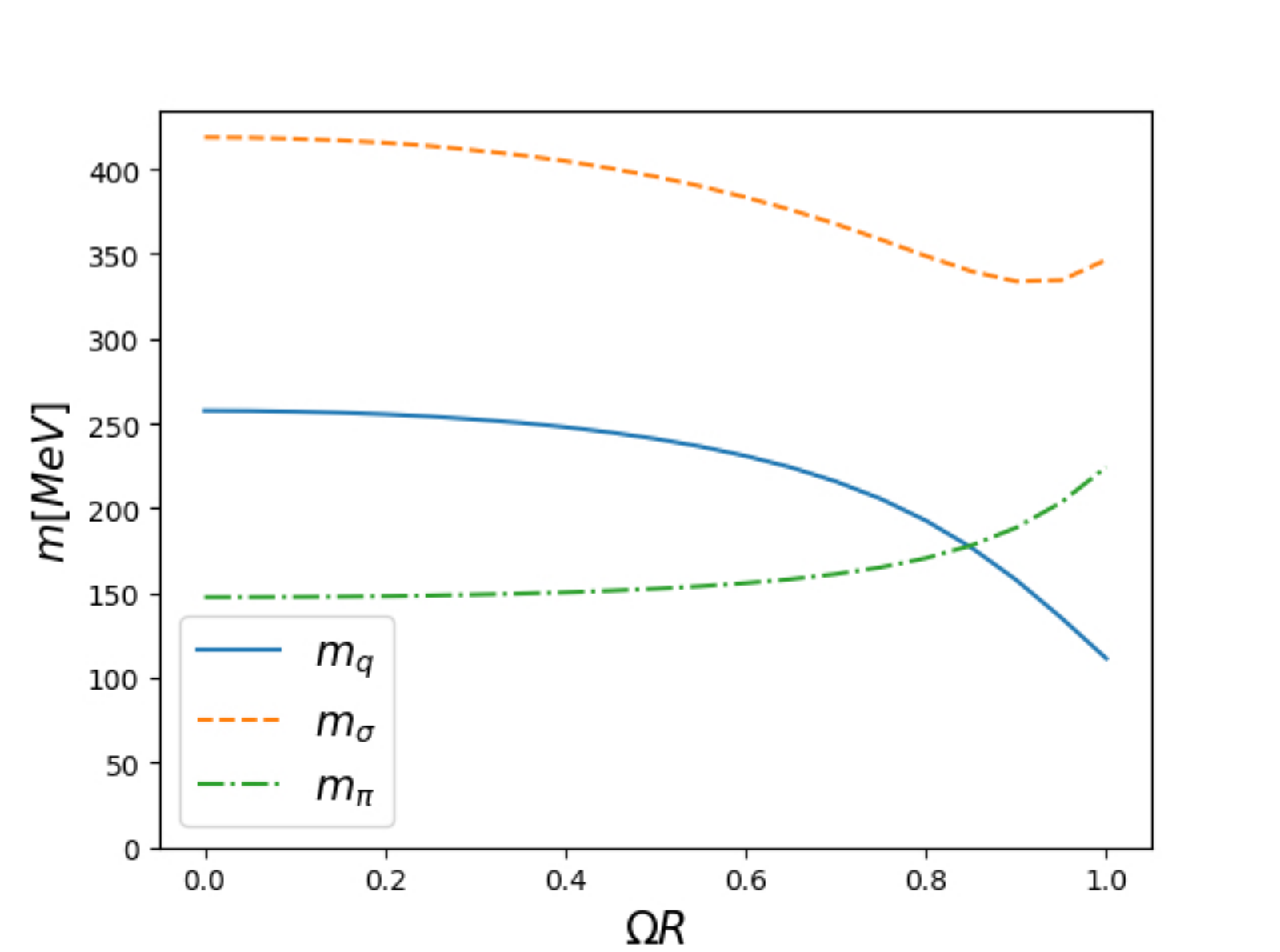}
    \caption{Meson masses and quark mass as functions of $\Omega$ at $T=120$ MeV from QM model.}
    \label{fig:grid T120}
\end{minipage}%
\hfill
\begin{minipage}[t]{0.45\linewidth}
\includegraphics[width=1\columnwidth]{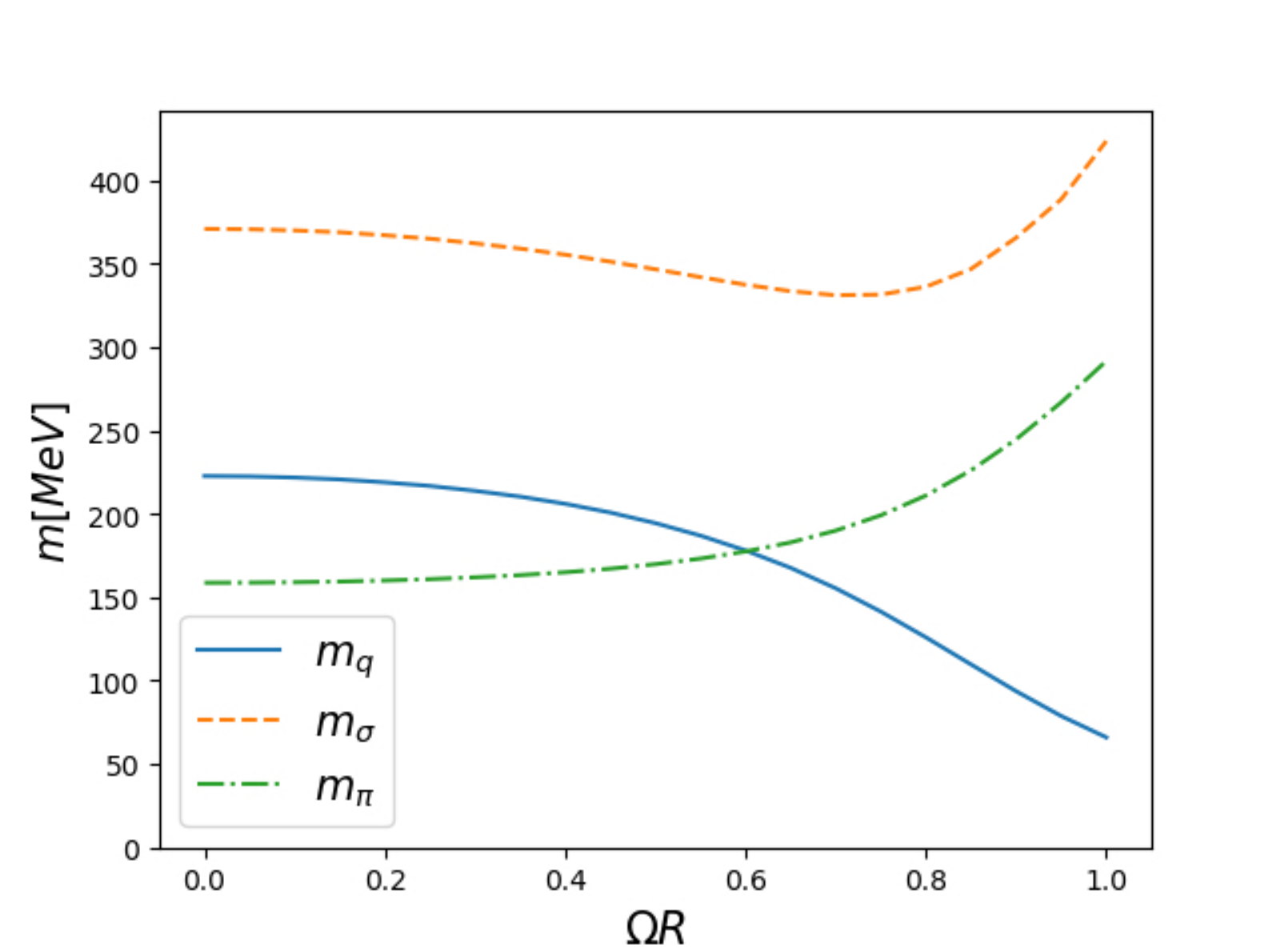}
    \caption{Meson masses and quark mass as functions of $\Omega$ at $T=140$ MeV from QM model.}
    \label{fig:grid T140}
\end{minipage}
\end{figure}

Quark mass and meson masses are presented in Fig.~\ref{fig:grid T120} and Fig.~\ref{fig:grid T140} for two different temperatures, respectively. In the finite-density case, $m_\s$ and $m_\pi$ will eventually be degenerate as the chemical potential increases~\cite{Tripolt:2013jra}. However, in our setup, we can only observe this tendency since the region $\O R> 1$ is inaccessible due to the causality restriction. We will see that the rotational effect is milder in our fRG calculation compared to the mean-field approximation (MFA) calculation that we will discuss later.

\begin{figure}[t]
\begin{minipage}[t]{0.45\linewidth}
\includegraphics[width=1\columnwidth]{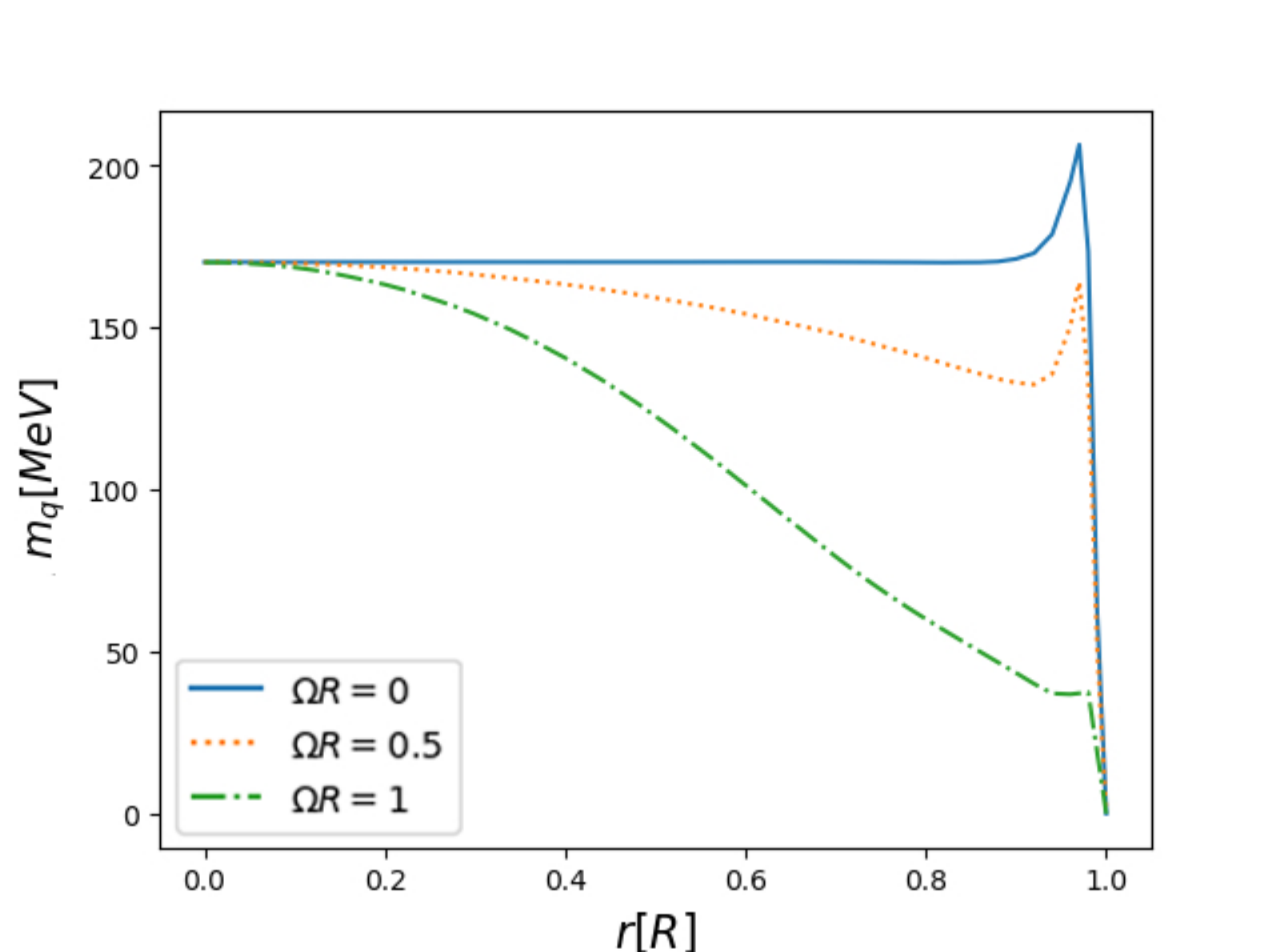}
    \caption{The quark mass as a function of the radius $r$ at different $\Omega$ at $T=160$ MeV.}
    \label{fig:mq_r}
\end{minipage}%
\hfill
\begin{minipage}[t]{0.45\linewidth}
\includegraphics[width=1\columnwidth]{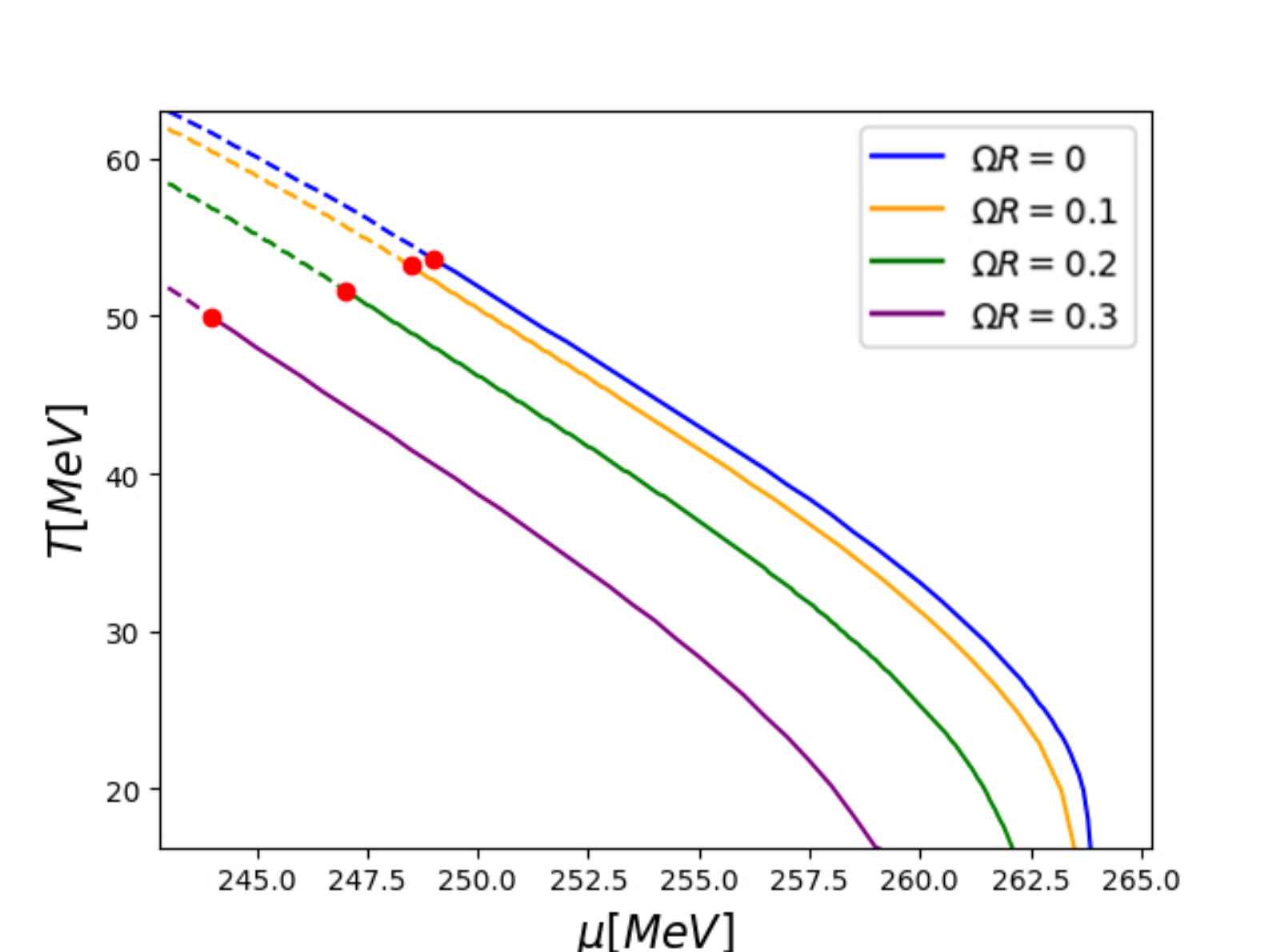}
    \caption{$T-\mu$ phase diagram near the critical end point at different angular velocity.}
    \label{fig:Tc-mu Omega}
\end{minipage}
\end{figure}

Figure.~\ref{fig:mq_r} displays the spatial dependence of quark mass at different angular velocities at $T=160$ MeV. It is apparent that the rotational suppression is strong near the boundary and becomes almost invisible at the center. This fact is easily understood: the modes with high angular momentum, which are located far from the center, experience a large effective chemical potential and are therefore more sensitive to rotation. Additionally, if we focus on the flow equation at the center ($r=0$), only the $l=0$ mode will contribute to the meson sector, and thus there is no rotational effect for mesons. However, fermions feel a constant ``baryon chemical potential'' of $\Omega/2$. Naively, one may expect a phase transition at large enough angular velocity at the center, similar to the finite density case. But with such a large $\Omega$, the system size should be very small, and boundary effects cannot be ignored. As a result, the causality constraint prevents sizable rotational effects at low temperatures. Note that the peak near the boundary is caused by finite-size effect, which is also observed in NJL model~\cite{Wang:2019nhd,Ebihara:2016fwa}.

Since it is hard to reach the critical end point (CEP) in the $T-\Omega$ phase diagram due to the causality constraint, we switch to the $T-\mu$ phase diagram to see how rotation affects the CEP. Here, $\mu$ is the quark chemical potential which is 1/3 of the baryon chemical potential. With our choice of parameters given by Eq.~(\ref{eq:para1}), the CEP of the $T-\mu$ diagram is located at a very low temperature of about $10$ MeV. In order to more clearly observe how rotation affects the CEP, we make another choice of parameters~\cite{Schaefer:2004en}:
\begin{equation}\label{eq:para2}
	\begin{split}
	\Lambda&=500\; \rm{MeV},\\
	m_\Lambda&=0,\\
	\lambda_\Lambda&=10,\\
	c&=0.
	\end{split}
\end{equation}
Note that with this choice of parameters, the phase transition line will split, and there will be another CEP at a low temperature~\cite{Schaefer:2004en}.  However, in this work, we only focus on the upper CEP. The numerical results at $r=0.9R$ are shown in Fig.~\ref{fig:Tc-mu Omega}, where dashed and solid lines represent second- and first-order phase transitions, respectively. We can see that the CEP slightly moves to lower temperature and lower density with increasing angular velocity, which can be expected from the analogy between rotation and chemical potential.

\section{Mean Field Approximation}
We can compare the above results with those obtained from mean field approximation (MFA) applied to the QM model. The thermodynamic potential under MFA reads
\begin{equation}
V_{\rm eff}=U(\sigma)-N_cN_f\int\frac{\ud p_z}{(2\pi)^2}T\sum_{l,i}\frac{1}{N^2_{l,i}}[\ln(1+\ue^{-\beta(\varepsilon-\Omega j)})+\ln(1+\ue^{-\beta(\varepsilon+\Omega j)})][J_l(\tilde p_{l,i}r)^2+J_{l+1}(\tilde p_{l,i}r)^2],
\end{equation}
where $\varepsilon=\sqrt{p_z^2+p_t^2+m^2}$ and the gap equation $V'_{\rm eff}(\sigma)=0$ reads
\begin{equation}
U'(\sigma)+g^2\sigma N_cN_f\int\frac{\ud p_z}{(2\pi)^2}\sum_{l,i}\frac{1}{N^2_{l,i}}\frac{1}{\varepsilon}[\frac{1}{1+\ue^{\beta(\varepsilon-\Omega j)}}+\frac{1}{1+\ue^{\beta(\varepsilon+\Omega j)}}][J_l(\tilde p_{l,i}r)^2+J_{l+1}(\tilde p_{l,i}r)^2]=0.
\end{equation}
The $\sigma$ and pion masses can be easily calculated by taking the second-order derivative of $V_{\rm eff}$.

\begin{figure}[t]
\begin{minipage}[t]{0.45\linewidth}
 \includegraphics[width=1\columnwidth]{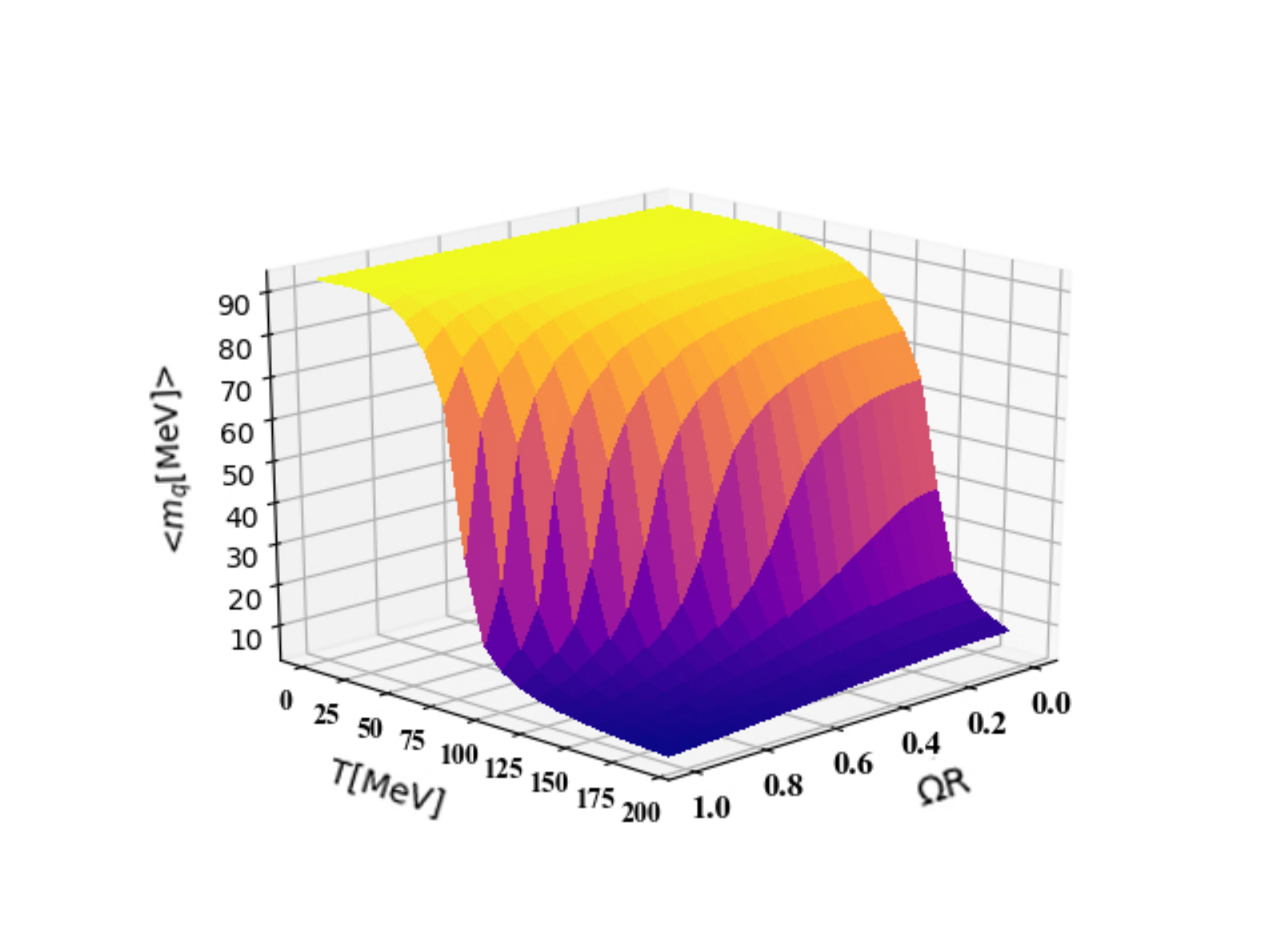}
    \caption{The quark mass $m_q$ as a function of $\Omega$ and $T$ at $r=0.9R$ under MFA of QM model.}
    \label{fig:MFA r09}
\end{minipage}%
\hfill
\begin{minipage}[t]{0.45\linewidth}
\includegraphics[width=1\columnwidth]{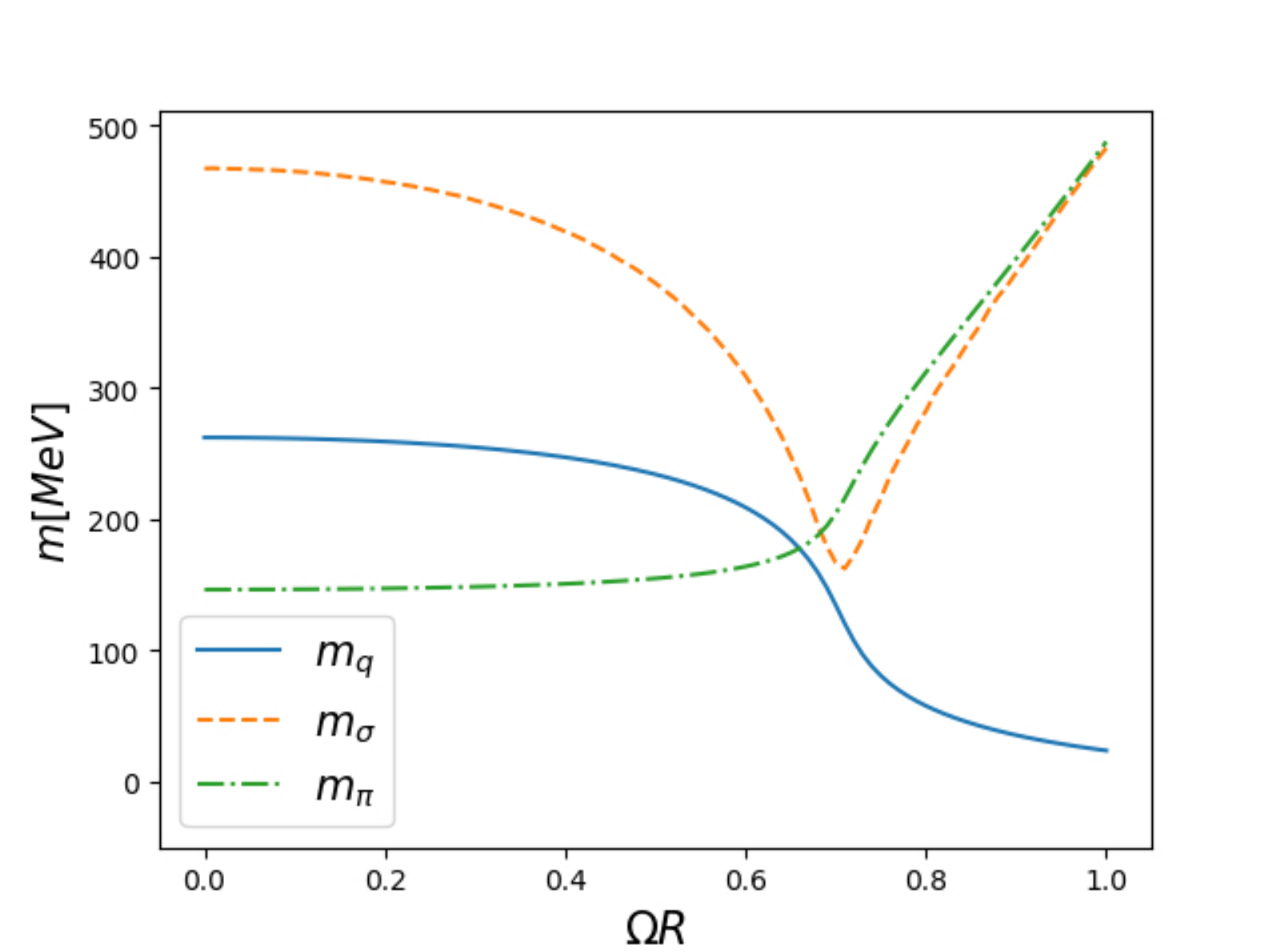}
    \caption{Meson masses and quark mass vs $\Omega$ at $T=120$ MeV and $r=0.9R$ under MFA of QM model.}
    \label{fig:MFA Omega}
\end{minipage}
\end{figure}

As shown in Figs.~\ref{fig:MFA r09} and \ref{fig:MFA Omega}, the rotational effect appears stronger with MFA than with the fRG method. In Fig.~\ref{fig:MFA Omega}, the pion mass and $\sigma$ mass nearly degenerate at large angular velocity, similar to the finite density case, while the fRG results (Fig.~\ref{fig:grid T120}) only show a tendency to degenerate. The pseudo-critical temperature lines from both methods are parallel, as illustrated in Fig.~\ref{fig:Tc_Omega}. Since MFA gives a lower $T_c$, it is expected that the pion mass and $\sigma$ mass will degenerate more easily than in the fRG case. However, due to the causality constraint, neither method can reach the critical end point, and the transition is always a crossover.

\section{Imaginary angular velocity}
Up to this point, we have considered real rotation. However, it is also academically interesting to consider imaginary rotation, i.e., a pure imaginary angular velocity, $\Omega \rightarrow i\Omega_I$. On the one hand, this bears similarities with the imaginary chemical potential case, which has been intensively studied and shown, for example, the interesting Roberge-Weiss (RW) periodicity and phase transition in QCD~\cite{Roberge:1986mm}. On the other hand, imaginary rotation is necessary for lattice simulations since real rotation leads to a sign problem in Monte-Carlo samplings~\cite{Yamamoto:2013zwa}.

\begin{figure}[t]
\begin{minipage}[t]{0.45\linewidth}
\includegraphics[width=1\columnwidth]{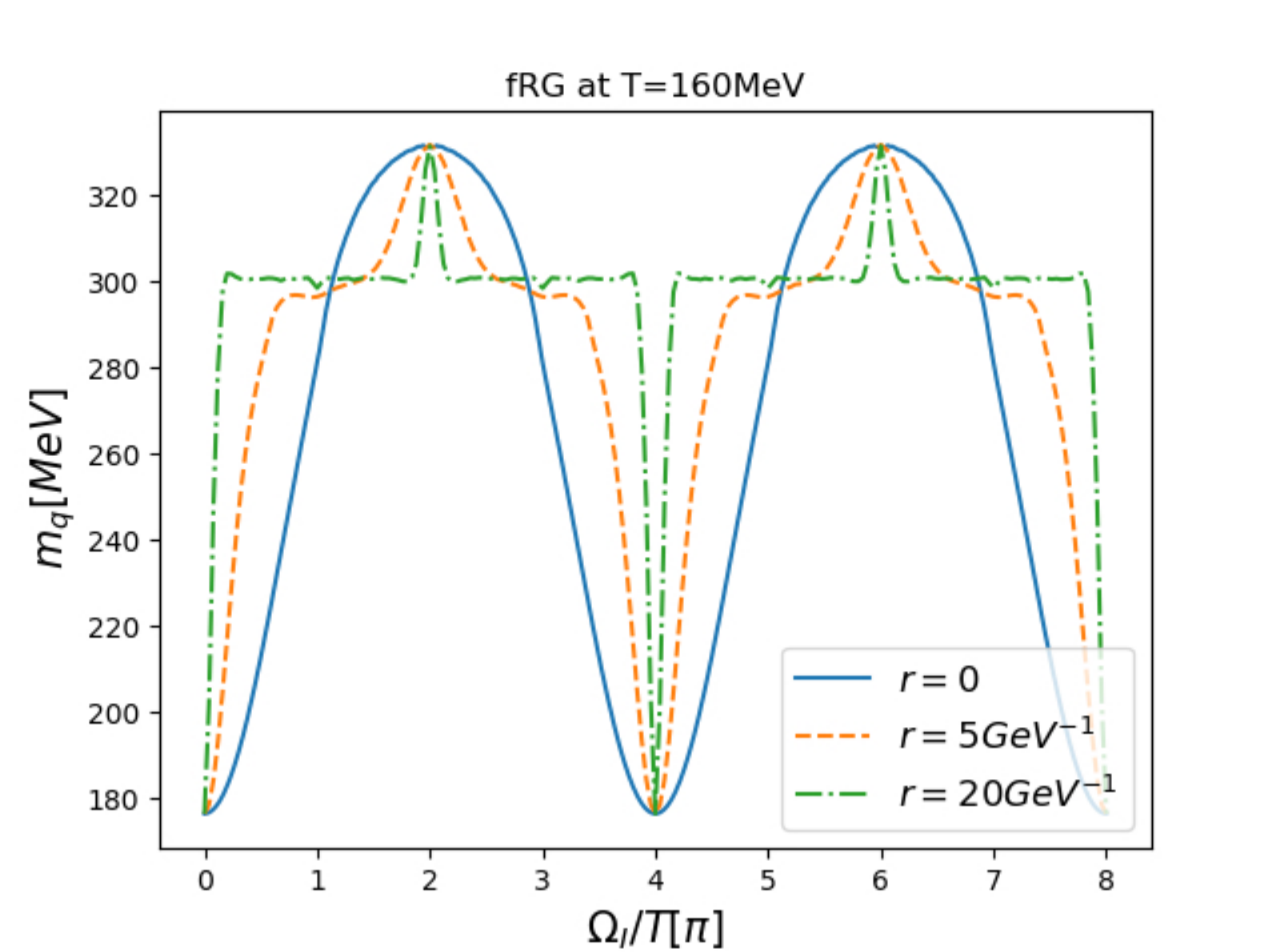}
    \caption{The quark mass as a function of the imaginary rotating angular velocity at $T=160$ MeV from fRG.}
    \label{fig:img_rot}
\end{minipage}%
\hfill
\begin{minipage}[t]{0.45\linewidth}
\includegraphics[width=1\columnwidth]{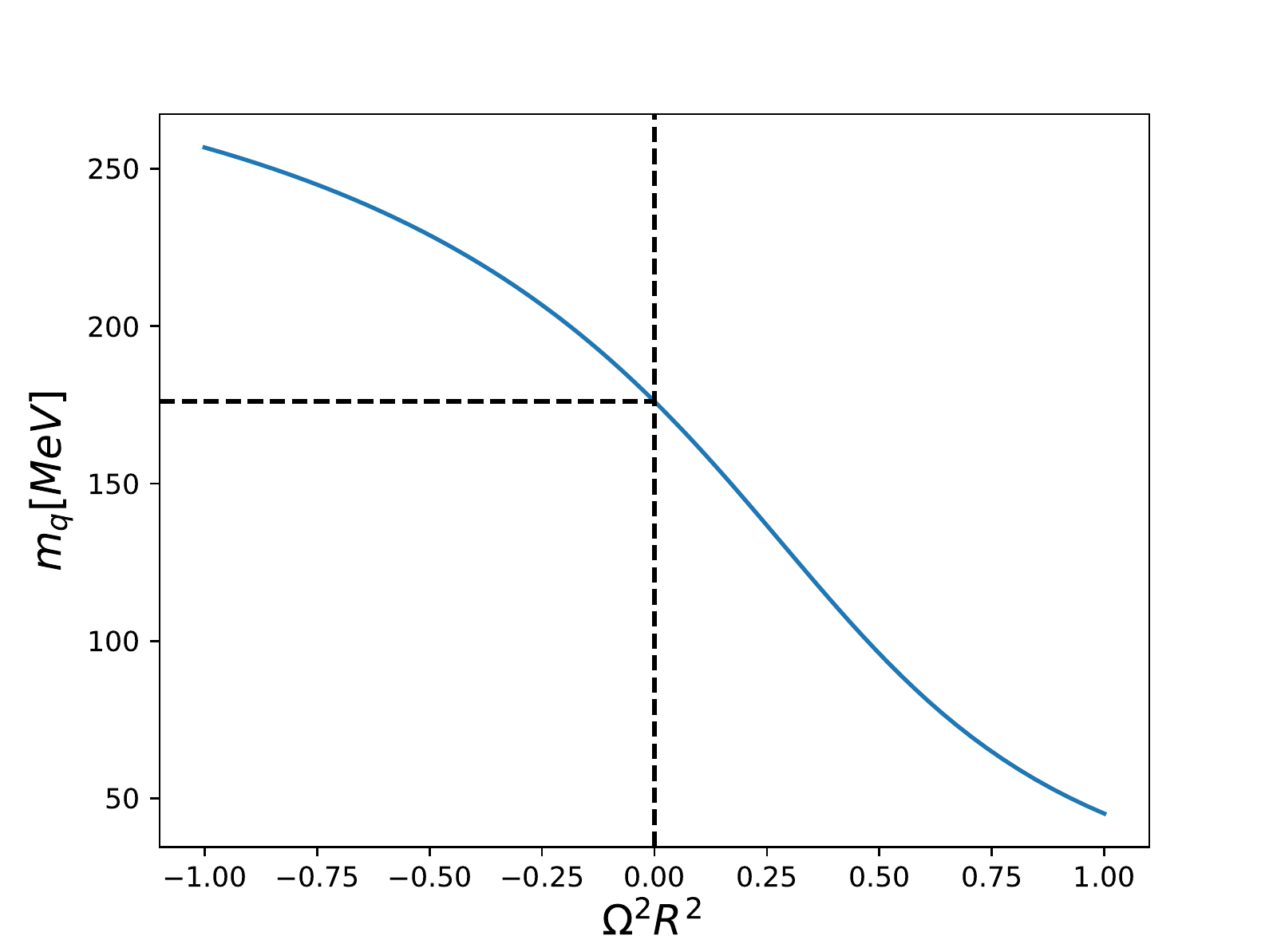}
    \caption{The quark mass as a function of the square of the rotating angular velocity at $T=160$ MeV and $r=0.9R$ from fRG calculation.}
    \label{fig:img_re_rot}
\end{minipage}%
\end{figure}

Under imaginary rotation, the fRG flow equation can be derived in a similar manner as under real rotation. The result is
\begin{equation}\label{eq:FlowEqRotImg}
\begin{split}
\partial_kU_k(r)=&\frac{1}{(2\pi)^2}\Big\{\sum_{l}{\tr}\frac{k}{2\varepsilon_\phi}\frac{\ue^{2\beta \varepsilon_\phi}-1}{\ue^{2\beta \varepsilon_\phi}+1-2\ue^{\beta \varepsilon_\phi}\cos \beta\Omega_I l}F_l(k,r)\\
&-N_cN_f\sum_{l}\frac{k}{\varepsilon_q}\frac{\ue^{2\beta \varepsilon_q}-1}{\ue^{2\beta \varepsilon_q}+1+2\ue^{\beta \varepsilon_q}\cos\beta \Omega_I j}[F_l(k,r)+F_{l+1}(k,r)]\Big\}.
\end{split}
\end{equation}
Similar to the case of imaginary chemical potential~\cite{Morita:2011jva}, the fRG flow equation remains real under imaginary rotation. Furthermore, since imaginary rotation does not break causality, a boundary condition is not necessary in this case, which enables integration over $p_t$ without the introduction of an infrared cutoff. The numerical result for quark mass is shown in Fig.~\ref{fig:img_rot}. Interestingly, under imaginary rotation, the chiral condensate is always enhanced, which is opposite to the behavior observed under real rotation.

As we do not include the contribution from the gauge field (Polyakov loop) in our model, we do not have the Roberge-Weiss (RW) phase transition~\cite{Roberge:1986mm} that is present in the case of imaginary baryon chemical potential. But a somehow trivial periodicity is observed in the chiral condensate. If we simply replace $\Omega_I$ with imaginary baryon chemical potential $\mu_I$, we would have a $2\pi$ period for $\mu_I/T$. In contrast, for the case of imaginary rotation, the periodicity is $4\pi$ for $\Omega_I/T$ due to the spin of the quark being $1/2$, which is twice that of the imaginary chemical potential case.

Focusing on the $[0,2\pi]$ region for $\Omega_I/T$, we observe that the chiral condensate at the center is a monotonic increasing function of the imaginary rotation. However, as we move away from the center, the quark mass increases very rapidly towards a certain value (the vacuum mass), and a small dip at $\Omega_I=\pi T$ appears. This behavior is more obvious in the NJL model, as illustrated in Fig.~\ref{fig:img_rot_NJL} ).

To provide a comprehensive analysis, we combine the results of imaginary and real angular velocities in a single plot, as shown in Fig.~\ref{fig:img_re_rot}. It becomes evident that the quark mass is a smooth function of the square of the angular velocity, which is a necessary condition for analytic continuation. Notably, a similar plot has been presented in lattice simulations for the case of baryon chemical potential (See, e.g,~\cite{Cea:2006yd}).

For the sake of comparison, we also present the results from (2+1)d NJL model within MFA under imaginary rotation in Appendix \ref{3dnjl}.

\section{Conclusion}
In this work, we calculate the chiral condensate using fRG of a rotating QM model. Our results show that rotation suppresses the chiral condensate at non-zero temperature, which agrees with a previous calculation using other models. We always take boundary conditions into account, either by directly imposing them or by introducing an effective $l$-dependent IR cutoff. In some cases, the critical end point in the $T-\Omega$ phase diagram may lie beyond the causality boundary (i.e., $\Omega R > 1$), and in the regions we have studied, the transition is always a crossover. We also confirm our fRG calculation by comparing it to a mean-field calculation of the QM model.

Furthermore, we study the effect of imaginary rotation using both the QM model and a (2+1)d NJL model. We find that the chiral condensate is a periodic function of $\Omega_I/T$. Away from the rotating center, the increase (or decrease) of the chiral condensate with respect to $\Omega_I$ becomes steeper, and small peaks appear at certain values of $\Omega_I$ ($\Omega_I/T=(2n+1)2\pi$ with $n$ integers).

It is important to note that in this work, we do not consider the contribution from gluons, which may play a crucial role in determining the phase transition line. However, incorporating gluon fields in a rotating spacetime within the fRG framework is a challenging task. It requires careful calculation of the propagators and interaction vertices of both gluons and ghosts to ensure causality. We expect that the boundary conditions may also affect the IR behavior of the gluon sector, which may influence the confinement-deconfinement behavior of QCD under rotation. We leave a detailed study of gluons in rotating spacetime to future work.

\section*{Acknowledgement}
We thank Kenji Fukushima for his collaboration and fruitful discussions in the early stages of this work. We also thank Guo-Liang Ma, Wei-Jie Fu and Ming-Hua Wei for helpful discussions. This work is supported by the National Key Research and Development Program of China (Grant No. 2022YFA1604900), the  Natural Science Foundation of China (Grant No.12247133,  No. 12225502 and No. 12075061), and the Natural Science Foundation of Shanghai (Grant No. 20ZR1404100). 

\bibliography{ref}

\appendix
\section{Unbounded system}\label{sec:unbound}
If we ignore the boundary, we have~\cite{Ebihara:2016fwa}
\begin{equation}
	\begin{split}
		p_{l,i}&\to p_t,\\
		\sum_i\frac{1}{N_{l,i}^2}&\to p_tdp_t,
	\end{split}\
\end{equation}
and the fRG flow equation becomes
\begin{equation}
\begin{split}
	\partial_kU_k=&\frac{1}{\beta V}\frac{1}{2}\int d^4x_E T\sum_n\int\frac{dp_z}{2\pi}\frac{p_tdp_t}{2\pi}\Big\{\sum_l{\tr}\frac{2k\theta(k^2-p^2)}{-(i\omega_n+\Omega l)^2+k^2+\frac{\partial^2U}{\partial\phi_i\partial\phi_j}}J_l(p_tr)^2\\
&-\sum_l2N_cN_f\frac{2k\theta(k^2-p^2)}{(\nu_n+i\Omega j)^2+k^2+g^2\phi^2}[J_l(p_tr)^2+J_{l+1}(p_tr)^2]\Big\}\\
=&\frac{1}{\beta V}\frac{1}{2}\int d^4x_E \int\frac{dp_z}{2\pi}\frac{p_tdp_t}{2\pi}\Big\{\sum_{i}{\tr}\frac{k\theta(k^2-p^2)}{\varepsilon_\phi}[1+n_B(\varepsilon_\phi+\Omega l)+n_B(\varepsilon_\phi-\Omega l)]J_l(p_{l,i}r)^2\\
&-\sum_{i}2N_cN_f\frac{k\theta(k^2-\tilde p^2)}{\varepsilon_q}[1-n_F(\varepsilon_q-\Omega j)-n_F(\varepsilon_q+\Omega j)][J_l(\tilde p_{l,i}r)^2+J_{l+1}(\tilde p_{l,i}r)^2]\Big\}.
\end{split}
\end{equation}

If we replace the rotation with quark chemical potential, i.e. $\Omega l\to 0$ and $\Omega j\to \mu$, we recover exactly the result in Ref.~\cite{Schaefer:2004en},
\begin{equation}\label{eq:EqTmu}
	\begin{split}
		\partial_kU_k=&T\sum_n\int\frac{d^3p}{(2\pi)^3}\Big\{{\tr}\frac{k\theta(k^2-p^2)}{-(i\omega_n)^2+k^2+\frac{\partial^2U}{\partial\phi_i\partial\phi_j}}-4N_cN_f\frac{k\theta(k^2-p^2)}{(\nu_n+i\mu)^2+k^2+g^2\phi^2}\Big\}\\
		=&\frac{k^4}{12\pi^2}\Big\{\frac{1}{\varepsilon_\sigma}\coth\frac{\beta\varepsilon_\sigma}{2}+\frac{3}{\varepsilon_\pi}\coth\frac{\beta\varepsilon_\pi}{2}-\frac{2N_cN_f}{\varepsilon_q}(\tanh\frac{\beta(\varepsilon_q-\mu)}{2}+\tanh\frac{\beta(\varepsilon_q+\mu)}{2})\Big\},
	\end{split}
\end{equation}
where we have used the identity
\begin{equation}
	\sum_{l=-\infty}^\infty J_l(x)^2=1.
\end{equation}

\section{(2+1)d NJL model }\label{3dnjl}
In this appendix, we study the (2+1)d NJL model under imaginary rotation. The (2+1)d NJL Lagrangian is
\begin{equation}
\Lag=\bar q[\gamma^0(\partial_\tau-i\Omega_I \hat J_z)-i\gamma^i\partial_i]	q+\frac{G}{2N_c}(\bar q q)^2.
\end{equation}
Under the mean field approximation (MFA) the effective potential reads
\begin{equation}
\begin{split}
\frac{V_{\rm eff}}{N_c}
=&V_0(m)-\int\frac{p_t\ud p_t}{2\pi}\sum_l \frac{1}{\b}\ln[1+\ue^{-2\b \varepsilon}+2\ue^{-\b \varepsilon}\cos (\b\Omega_I j)][J_l^2(p_t r)+J_{l+1}^2(p_t r)],
\end{split}
\end{equation}
where $m$ is the dynamic quark mass, $\varepsilon=\sqrt{p_z^2+p_t^2+m^2}$ and $V_0(m)$ is the vacuum ($T=\O=0$) contribution
\begin{equation}
\begin{split}
V_{0}
=&\frac{m^2}{2}(\frac{1}{G}-\frac{\L}{\p})+\frac{m^3}{3\p}.
\end{split}
\end{equation}
To arrive at this expression we have assumed $\L \gg m$ and droped the $m$ independent terms. We can define the dynamical mass of quarks in the vacuum $m_0$ by using the gap equation in vacuum
\begin{equation}
\frac{1}{G(\L)}=\frac{\L-m_0}{\p},
\end{equation}
then it is easy to verify that $m_0$ is the only free model parameter of the (2+1)d NJL model. The vacuum effective potential becomes
\begin{equation}
\begin{split}
V_{0}=&\frac{m^3}{3\p}-\frac{m^2m_0}{2\p}.
\end{split}
\end{equation}
Then the gap equation at finite temperature and imaginary angular velocity is
\begin{equation}\label{eq:gap_njl}
\begin{split}
m(m-m_0)+\int p_t\ud p_t\sum_l \frac{m}{2\varepsilon}(\frac{1}{\ue^{\b(\varepsilon+i\Omega_Ij)}+1}+\frac{1}{\ue^{\b(\varepsilon-i\Omega_Ij)}+1})[J_l^2(p_t r)+J_{l+1}^2(p_t r)]=0.
\end{split}
\end{equation}
If we expand the fermion distribution function as $1/[\ue^{\b(\varepsilon\pm i\Omega_Ij)}+1]=-\sum_{n=1}^\infty(-1)^n\ue^{-n\b(\varepsilon\pm i\Omega_Ij)}$, then the integration over $p_t$ and summation over $l$ can be performed exactly by using the identities
\begin{equation}
\begin{split}
&\sum_{n\in \mathbb{Z}}\ue^{in\theta}J^2_n(x)=J_0(x\sqrt{2-2\cos\theta}),\\
&\int_1^\infty\ue^{-ax}J_0(b\sqrt{x^2-1})\ud x=\frac{1}{\sqrt{a^2+b^2}}\ue^{-\sqrt{a^2+b^2}}.
\end{split}
\end{equation}
Finally, we arrive at
\begin{equation}\label{eq:gap_njl_n}
\begin{split}
m(m-m_0)-\sum_{n\neq 0}(-1)^n\frac{m\ue^{-m\sqrt{(n\b)^2+2r^2[1-\cos (n\b\O_I)]}}}{\sqrt{(n\b)^2+2r^2[1-\cos (n\b\O_I)]}}\cos (n\b\frac{\O_I}{2})=m(m-m_0)-\tr(S_\b-S_0)=0,
\end{split}
\end{equation}
where $S_0$ represents the fermion propagator in vacuum, while the thermal propagator $S_\b(\D\t)$ is defined as $\sum_n S_0(\D\t+n\b)$. This propagator satisfies the periodic boundary condition $S_\b(\D\t)=S_\b(\D\t+\b)$. The physical meaning of $n$ can be understood as the winding number along the compacted time direction in Euclidean spacetime. Here, $n=0$ corresponds to the vacuum contribution, which is just the $m_0$ term. The contribution from fermion spin in this expression is $(-1)^n\cos(n\b{\O_I}/{2})$, while the remaining part in the summation is purely bosonic.

It should be noted that Eq.(\ref{eq:gap_njl_n}) is only valid when dealing with imaginary rotation. Naive analytic continuation will lead to singularity in the summation. When the angular velocity is real, the metric becomes complex in Euclidean spacetime, rendering the winding prescription inapplicable. A similar situation was also observed in\cite{Chen:2022smf} when calculating the perturbative Polyakov-loop potential.

\begin{figure}
\begin{minipage}[t]{0.45\linewidth}
	\includegraphics[width=1\columnwidth]{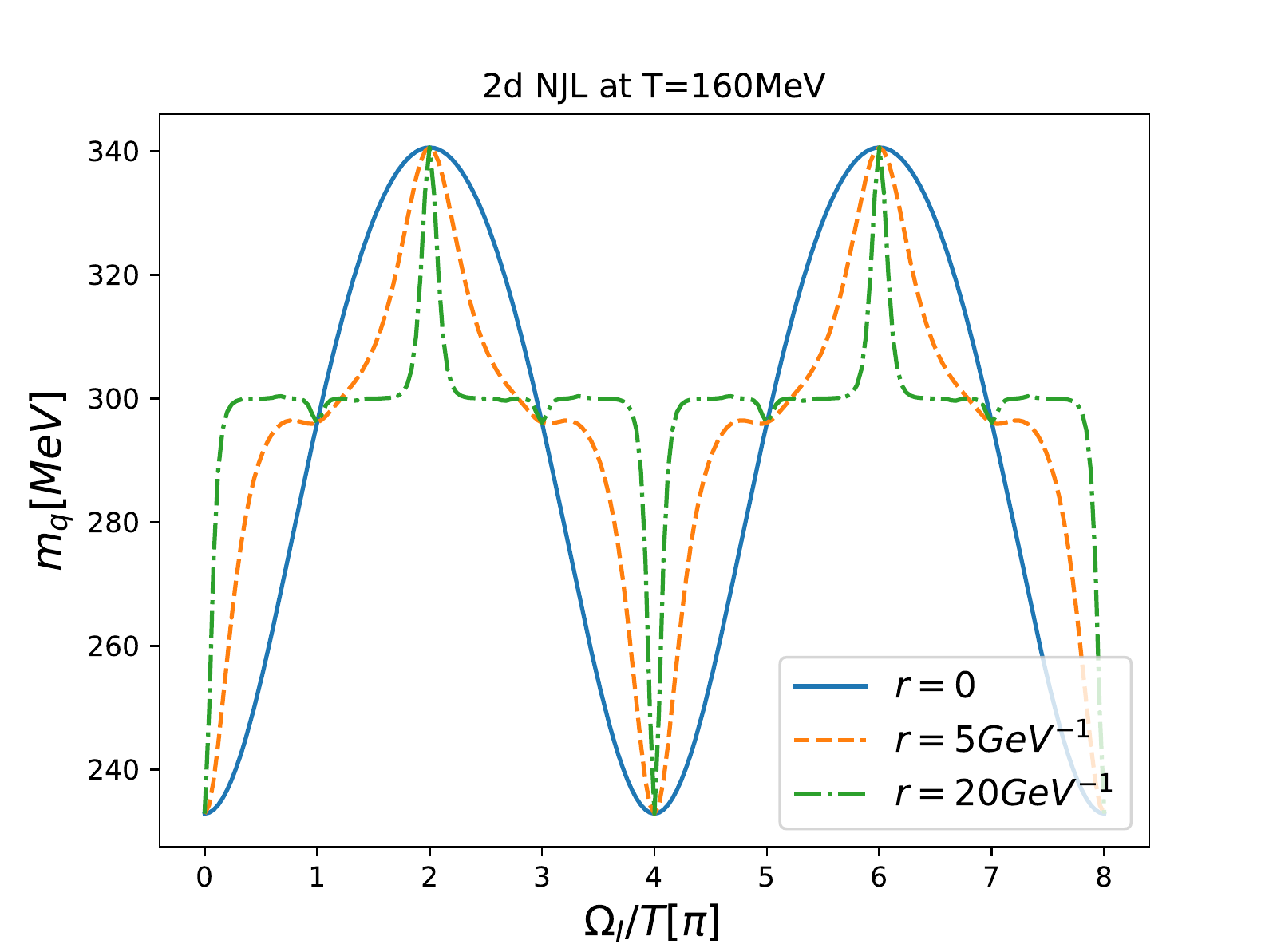}
		\caption{The quark mass as a function of the imaginary angular velocity at $T=160$ MeV from (2+1)d NJL model, with parameter $m_0=300$ MeV.}
		\label{fig:img_rot_NJL}
	\end{minipage}
\hfill
\begin{minipage}[t]{0.45\linewidth}
\includegraphics[width=1\columnwidth]{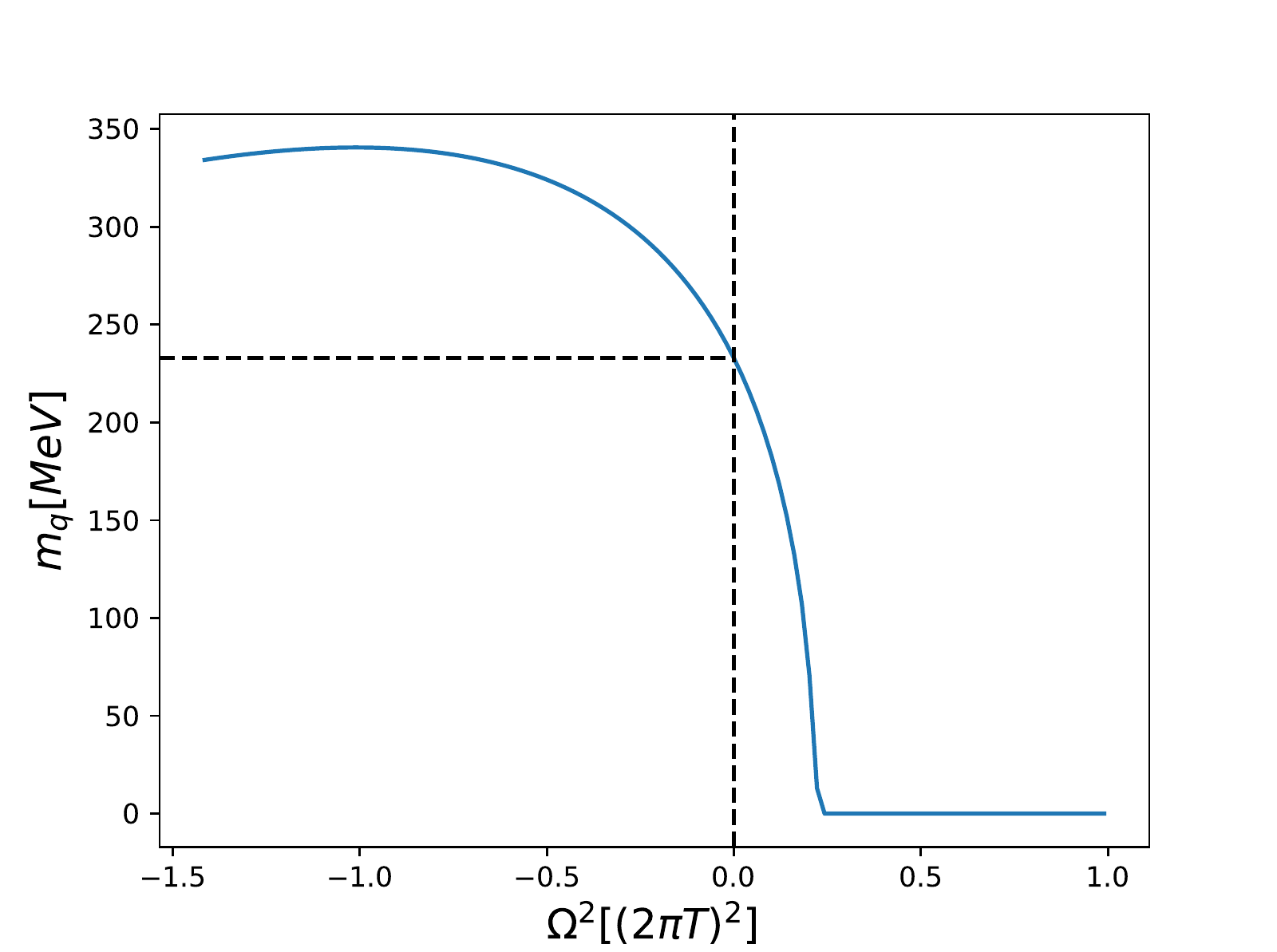}
    \caption{The quark mass as a function of  the square of rotating angular velocity at $T=160$ MeV and $r=0$ with 2d NJL model.}
    \label{fig:img_re_rot_njl}
\end{minipage}%
	\end{figure}

The numerical solution to Eq.(\ref{eq:gap_njl_n}) can be easily obtained, as the summation over $n$ converges very rapidly. The resulting Fig.~\ref{fig:img_rot_NJL} is in good agreement with the fRG Fig.~\ref{fig:img_rot}. At $\beta \Omega_I=\pi$, it is easy to verify that Eq.~(\ref{eq:gap_njl}) or Eq.~(\ref{eq:gap_njl_n}) become independent of $r$, which is not the case in the QM model Eq.~(\ref{eq:FlowEqRotImg}) due to the meson contribution.

We give a plot of quark mass as a function of the square of angular velocity in Fig.~(\ref{fig:img_re_rot_njl}). Since we only have fermionic degrees of freedom in NJL model, it is possible to go beyond the causality region without encountering singularity. Interestingly, as we observe in this plot, the quark mass gradually diminishes to zero as we increase $\Omega^2$ into the $\Omega R > 1$ region. 

\end{document}